\documentclass[conference]{IEEEtran}
\IEEEoverridecommandlockouts
\usepackage{bbm}

\usepackage{cite}
\usepackage{amsmath,amssymb,amsfonts}
\usepackage{graphicx}
\usepackage{textcomp}
\usepackage{xcolor}
\usepackage{cite}
\usepackage{dblfloatfix}
\usepackage{subcaption}
\usepackage{overpic}
\usepackage{amsmath,amssymb,amsfonts}

\usepackage{graphicx}
\usepackage{textcomp}
\usepackage{xcolor}
\usepackage{float}
\usepackage{amsthm}
\usepackage{graphicx}
\usepackage{epstopdf}
\usepackage{amsmath,bm}
\usepackage{amsfonts}
\usepackage{amssymb}
\usepackage{color}
\usepackage{multirow}
\usepackage{multicol}
\usepackage{soul,xcolor}
\usepackage{algorithm}
\usepackage{algpseudocode}

\usepackage{comment}

\theoremstyle{plain}

\newcommand{\vect}[1]{\mathbf{#1}}

\def\diag{\mathrm{diag}}

\def\rank{\mathrm{rank}}
\def\Htran{\mbox{\tiny $\mathrm{H}$}}
\def\Ttran{\mbox{\tiny $\mathrm{T}$}}

\def\BibTeX{{\rm B\kern-.05em{\sc i\kern-.025em b}\kern-.08em
    T\kern-.1667em\lower.7ex\hbox{E}\kern-.125emX}}
\begin{document}

\title{Fronthaul-Efficient Cell-Free Massive MIMO via Stream-Adaptive Resolution Control \\
\thanks{This work was carried out within the scope of the project 122C149 – Intelligent End-to-End Design of Energy-Efficient and Hardware Impairments-Aware Cell-Free Massive MIMO for Beyond 5G. \"O. T. Demir was supported by the 2232-B International Fellowship for Early Stage Researchers Programme funded by the Scientific and Technological Research Council of Türkiye (TÜBİTAK).}
}

\author{\IEEEauthorblockN{Özlem Tuğfe Demir}
\IEEEauthorblockA{\textit{Department of Electrical and Electronics Engineering} \\
\textit{Bilkent University}\\
Ankara, Turkiye \\
E-mail: ozlemtugfedemir@bilkent.edu.tr}\and
\IEEEauthorblockN{Sinan Gezici}
\IEEEauthorblockA{\textit{Department of Electrical and Electronics Engineering} \\
\textit{Bilkent University}\\
Ankara, Turkiye \\
E-mail: gezici@ee.bilkent.edu.tr}
\vspace{-6mm}}

\maketitle

\begin{abstract}
We consider the uplink of a fronthaul-constrained cell-free massive MIMO system with a single multi-antenna user equipment (UE) transmitting multiple spatial streams to distributed access points (APs). Due to heterogeneous channel conditions and limited fronthaul capacity, uniform-resolution quantization across APs and streams is highly inefficient. To address this, we propose a stream- and AP-adaptive resolution control framework that jointly optimizes the transmit power and per-AP per-stream quantization resolutions. By leveraging a distributed implementation of singular-value decomposition (SVD)-based processing, the data streams are decoupled at the APs, enabling flexible fronthaul compression across both spatial streams and APs. The resulting non-convex optimization problem is tackled using a weighted minimum mean-squared error (WMMSE)-based block coordinate descent algorithm with semi-closed-form updates.
\end{abstract}

\section{Introduction}

Cell-free massive multiple-input multiple-output (MIMO) has emerged as a key technology for beyond-5G and 6G systems due to its ability to provide uniformly high data rates, particularly for user equipments (UEs) experiencing unfavorable propagation conditions in conventional co-located cellular architectures. By distributing a large number of access points (APs) over a wide area and jointly serving UEs, cell-free massive MIMO eliminates cell boundaries and significantly improves coverage and spectral efficiency.

In the uplink of a cell-free massive MIMO system, the APs perform local signal processing and forward the processed signals to a central processing unit (CPU) for final data detection. This architecture closely aligns with modern Open RAN implementations, where the MIMO processing is functionally split between the radio units and the baseband unit. However, the fronthaul links connecting these units are capacity-limited, which introduces additional quantization distortion when the locally processed signals are forwarded using finite-resolution representations \cite{khorsandmanesh2023optimized,masoumi2019performance}.

The impact of fronthaul quantization on MIMO systems has been recently investigated in the point-to-point setting, where it has been shown that jointly optimizing the transmit power and the quantization resolution across data streams yields substantial performance gains over conventional schemes \cite{stream-adaptive}. Despite these advances, the existing literature has primarily focused on co-located MIMO systems or cell-free setups with single-antenna UEs \cite{kim2024meta}. In contrast, the interplay between spatial multiplexing and fronthaul quantization in cell-free architectures with multi-antenna UEs remains largely unexplored. 

In this paper, we aim to fill this gap by extending the stream-adaptive quantization framework to cell-free massive MIMO systems with a multi-antenna UE. By leveraging a distributed implementation of the classical singular-value decomposition (SVD)-based MIMO processing, we decouple the data streams at the APs and enable per-stream quantization over the fronthaul. Building on this structure, we propose a novel joint optimization framework that adapts both the transmit power and the quantization resolution across APs and streams under a total fronthaul constraint.

\section{System Model}

We consider the uplink of a cell-free massive MIMO system where a single UE equipped with $K$ antennas communicates with a CPU through $L$ distributed APs. Each AP is equipped with $N$ antennas and is connected to the CPU via fronthaul links with limited capacity.

\subsection{Uplink Signal Model}

We let $\mathbf{H}_l\in\mathbb{C}^{N\times K}$ denote the channel matrix between the UE and AP $l$. By stacking the channels from all the APs, we define the aggregate channel matrix as
\begin{align}
    \mathbf{H}=
    \begin{bmatrix}
        \mathbf{H}_1^{\Ttran} & \cdots & \mathbf{H}_L^{\Ttran}
    \end{bmatrix}^{\Ttran}
    \in\mathbb{C}^{LN\times K}.
\end{align}
The UE transmits up to $d=\rank(\mathbf{H})$ independent data streams collected in the vector $\mathbf{s}\sim\mathcal{CN}(\mathbf{0},\mathbf{P})$, where $\mathbf{P}=\diag(p_1,\ldots,p_d)$. Later, if the quantization bit resolution of a stream is zero or the allocated transmit power is zero, then that stream will be deactivated. We let $\mathbf{F}\in\mathbb{C}^{K\times d}$ denote the precoder; hence, the transmitted signal becomes $\mathbf{x}=\mathbf{F}\mathbf{s}$. Then, the received signal at AP $l$ is given by 
\vspace{-0.1cm}\begin{align}
    \mathbf{y}_l=\mathbf{H}_l\mathbf{x}+\mathbf{n}_l
    =\mathbf{H}_l\mathbf{F}\mathbf{s}+\mathbf{n}_l,
\end{align}

\vspace{-0.1cm}

\noindent for $l\in\{1,\ldots,L\}$, where $\mathbf{n}_l\sim\mathcal{CN}(\mathbf{0},\sigma^2\mathbf{I}_N)$ is the additive complex Gaussian noise with independent components. 

\subsection{Quantization-Free Optimal Processing}

To reveal the optimal structure, we first omit fronthaul quantization, and define the aggregate received signal and noise as $
    \mathbf{y}=
    \begin{bmatrix}
        \mathbf{y}_1^{\Ttran} & \cdots & \mathbf{y}_L^{\Ttran}
    \end{bmatrix}^{\Ttran}\in\mathbb{C}^{LN},
$
and
$
    \mathbf{n}=
    \begin{bmatrix}
        \mathbf{n}_1^{\Ttran} & \cdots & \mathbf{n}_L^{\Ttran}
    \end{bmatrix}^{\Ttran}\in\mathbb{C}^{LN}$, respectively. Hence, the concatenated received signal is given by
\vspace{-0.1cm}\begin{align}   \mathbf{y}=\mathbf{H}\mathbf{F}\mathbf{s}+\mathbf{n}.
\end{align}

We express the compact SVD of $\mathbf{H}$ as
\begin{align}    \mathbf{H}=\mathbf{U}\mathbf{\Lambda}\mathbf{V}^{\Htran},
\end{align}
where $\mathbf{U}\in\mathbb{C}^{LN\times d}$ and $\mathbf{V}\in\mathbb{C}^{K\times d}$ are semi-unitary, and $\mathbf{\Lambda}=\diag(\lambda_1,\ldots,\lambda_d)$ contains the non-zero singular values in descending order.

Following the capacity-achieving structure of point-to-point MIMO channels, we choose $ \mathbf{F}=\mathbf{V}$. Then, applying $\mathbf{U}^{\Htran}$ at the CPU yields
\vspace{-0.1cm}\begin{align}
    \tilde{\mathbf{y}}
    =\mathbf{U}^{\Htran}\mathbf{y}
    =\mathbf{\Lambda}\mathbf{s}+\tilde{\mathbf{n}},
\end{align}
where $\tilde{\mathbf{n}}=\mathbf{U}^{\Htran}\mathbf{n}\sim\mathcal{CN}(\mathbf{0},\sigma^2\mathbf{I}_d)$.

Hence, in the absence of fronthaul quantization, the cell-free uplink is diagonalized into $d$ parallel streams.

\subsection{Distributed Structure of the Global Combiner}

Although the optimal combiner $\mathbf{U}^{\Htran}$ is defined centrally, it can be decomposed across the APs. In particular, we partition
\begin{align}
    \mathbf{U}=
    \begin{bmatrix}
        \mathbf{U}_1^{\Ttran} & \cdots & \mathbf{U}_L^{\Ttran}
    \end{bmatrix}^{\Ttran},
\end{align}
where each $\mathbf{U}_l\in\mathbb{C}^{N\times d}$ corresponds to AP $l$. Then,
\vspace{-0.1cm}\begin{align}
    \tilde{\mathbf{y}}
    =\mathbf{U}^{\Htran}\mathbf{y}
    =\sum_{l=1}^{L}\mathbf{U}_l^{\Htran}\mathbf{y}_l.
\end{align}
Hence, each AP computes
\vspace{-0.1cm}\begin{align}
    \mathbf{r}_l=\mathbf{U}_l^{\Htran}\mathbf{y}_l
 =\mathbf{U}_l^{\Htran}\mathbf{H}_l\mathbf{V}\mathbf{s}
    +\mathbf{U}_l^{\Htran}\mathbf{n}_l.
\end{align}

The locally processed signals $\mathbf{r}_l$ are then forwarded from each AP to the CPU over the fronthaul links. At the CPU, the received vectors are coherently combined by simple summation: $\vect{r} = \sum_{l=1}^{L} \mathbf{r}_l$.

\subsection{Per-AP Per-Stream Quantized Fronthaul}

In the case of limited-capacity fronthaul links, each AP quantizes every processed stream separately before fronthaul transmission. We let $r_{l,i}$ denote the $i$th entry of $\mathbf{r}_l$. Using the Bussgang decomposition, the quantized signal is modeled as
\begin{align}
    \hat{r}_{l,i}=(1-\beta_{l,i})r_{l,i}+q_{l,i},
\end{align}
where $\beta_{l,i}\in(0,1]$ is the distortion factor and $q_{l,i}$ is the quantization noise. We define
\begin{align}
    \mathbf{B}_l=\diag(\beta_{l,1},\ldots,\beta_{l,d}).
\end{align}
Then,
\begin{align}
    \hat{\mathbf{r}}_l=(\mathbf{I}_d-\mathbf{B}_l)\mathbf{r}_l+\mathbf{q}_l.
\end{align}
By construction, $\mathbf{q}_l$ is uncorrelated with $\mathbf{r}_l$. Moreover, the covariance matrix of $\mathbf{q}_l$ is, in general, non-diagonal due to the correlation among the entries of $\mathbf{r}_l$ \cite{demir2020bussgang}. The $i$th diagonal element of the covariance matrix is given by
\begin{align}
    \mathbb{E}\left\{|q_{l,i}|^2\right\}
    =
    \beta_{l,i}(1-\beta_{l,i})
    \mathbb{E}\left\{|r_{l,i}|^2\right\}.
\end{align}

At the CPU, the attenuation introduced by the Bussgang decomposition is compensated by normalizing each quantized branch as
\begin{align}
    \tilde{r}_{l,i}
    \triangleq
    \frac{\hat{r}_{l,i}}{1-\beta_{l,i}}
    =
    r_{l,i}+\frac{q_{l,i}}{1-\beta_{l,i}}, \label{eq:tilderli}
\end{align}
for all AP-stream branches that are assigned a non-zero quantization resolution. If no bit is assigned to a given AP-stream branch, then that branch is inactive and no signal is forwarded from that branch to the CPU.

The CPU then coherently combines the normalized signals corresponding to the same streams: $\tilde{\mathbf{r}}=\sum_{l=1}^{L}\tilde{\mathbf{r}}_l$. Hence, the equivalent input-output relation can be written as
\begin{align}
    \tilde{\mathbf{r}}=\bar{\mathbf{G}}\mathbf{s}+\bar{\mathbf{z}},
\end{align}
where $\bar{\mathbf{G}}
    =
    \sum_{l=1}^{L}\mathbf{A}_l\mathbf{U}_l^{\Htran}\mathbf{H}_l\mathbf{V}$
and
$ \bar{\mathbf{z}}
    =
    \sum_{l=1}^{L}
    \mathbf{A}_l
    \left(
    \mathbf{U}_l^{\Htran}\mathbf{n}_l
    +
    (\mathbf{I}_d-\mathbf{B}_l)^{-1}\mathbf{q}_l
    \right)$.
Here, $\mathbf{A}_l=\diag(a_{l,1},\ldots,a_{l,d})$ is a diagonal activity matrix with
\begin{align}
    a_{l,i}=
    \begin{cases}
        1, & \text{if the $(l,i)$th AP-stream branch is active},\\
        0, & \text{otherwise}.
    \end{cases}
\end{align}

\subsection{Exact Achievable Rate}

Due to the coherent combining of the normalized quantized signals across the APs, the effective disturbance $\bar{\mathbf{z}}$ contains contributions from both the receiver noise and the normalized quantization noise. Since the same data streams are processed and quantized across different APs, the resulting quantization noise components are, in general, correlated. In addition, the local combiners $\mathbf{U}_l$ are not individually semi-unitary, which also contributes to a colored effective disturbance.

We let $\bar{\mathbf{R}}_z=\mathbb{E}\{\bar{\mathbf{z}}\bar{\mathbf{z}}^{\Htran}\}$ denote the covariance matrix of the effective disturbance. By treating $\bar{\mathbf{z}}$ as the worst-case Gaussian noise, an achievable uplink rate is given by
\begin{align}
    R_{\rm exact}
    =
    \log_2\det\!\left(
    \mathbf{I}_d+\bar{\mathbf{G}}\mathbf{P}\bar{\mathbf{G}}^{\Htran}\bar{\mathbf{R}}_z^{-1}
    \right).
    \label{eq:Rexact}
\end{align}
In the numerical evaluations, the covariance matrix $\bar{\mathbf{R}}_z$ is estimated via Monte-Carlo simulations using the actual Lloyd--Max quantizers applied to the AP-stream branches.

\subsection{Optimization-Oriented Approximation}

The exact rate expression in \eqref{eq:Rexact} involves a colored noise covariance matrix, the closed-form expression of which is in general not available due to the correlation induced by the quantization of coupled signals across the APs. This makes the joint optimization of the transmit powers and quantization parameters intractable. To obtain a tractable formulation, we introduce the following approximations.

For optimization purposes, we assume that the quantization noise components are uncorrelated across APs and streams. In addition, during the continuous optimization phase, each AP--stream pair is assumed to be assigned at least a non-zero quantization resolution. After compensating for the Bussgang attenuation via normalization, the coherently combined signal preserves the diagonal structure induced by the global SVD. Under this approximation, the streams become effectively decoupled, which enables a tractable per-stream signal-to-interference-plus-noise ratio (SINR) expression.

It is important to emphasize that these approximations are used only for optimization purposes. The exact rate expression in \eqref{eq:Rexact}, evaluated numerically via Monte-Carlo simulations with the full covariance matrix, is used for performance evaluation. 

Since the quantization-free uplink is diagonalized by the global SVD, the desired signal power of stream $i$ after the normalization step is approximated as $ p_i\lambda_i^2$, where $\lambda_i$ is the $i$th singular value of the aggregate channel matrix. Moreover, we define
\vspace{-2mm}
\begin{align}
    \rho_{l,i}
    \triangleq
    \mathbb{E}\left\{|r_{l,i}|^2\right\}
    =
    \sum_{k=1}^{d}
    p_k
    \left|
    \mathbf{u}_{l,i}^{\Htran}\mathbf{H}_l\mathbf{v}_k
    \right|^2
    +
    \sigma^2\|\mathbf{u}_{l,i}\|^2,
\end{align}
where $\mathbf{u}_{l,i}$ is the $i$th column of $\mathbf{U}_l$ and $\mathbf{v}_k$ is the $k$th column of $\mathbf{V}$.

By the Bussgang decomposition, the variance of the normalized quantization noise becomes
\vspace{-2mm}
\begin{align}
    \mathbb{E}\left\{
    \left|
    \frac{q_{l,i}}{1-\beta_{l,i}}
    \right|^2
    \right\}
    =
    \frac{\beta_{l,i}}{1-\beta_{l,i}}\,\rho_{l,i}.
\end{align}
Treating the quantization distortions from different APs and streams as independent, the effective disturbance variance of stream $i$ is approximated as
\vspace{-2mm}
\begin{align}
    \nu_i
    =
    \sigma^2
    +
    \sum_{l=1}^{L}
    \frac{\beta_{l,i}}{1-\beta_{l,i}}\rho_{l,i}. \label{eq:nui}
\end{align}
Accordingly, the approximate per-stream SINR is written as
\begin{align}
    \bar{\gamma}_i
    =
    \frac{p_i\lambda_i^2}{\nu_i}
    =
    \frac{p_i\lambda_i^2}{
    \sigma^2+\sum_{l=1}^{L}\frac{\beta_{l,i}}{1-\beta_{l,i}}\rho_{l,i}
    }.
\end{align}
Hence, the corresponding surrogate sum rate is given by
\begin{align}
    \bar{R}
    =
    \sum_{i=1}^{d}
    \log_2\left(
    1+\frac{p_i\lambda_i^2}{
    \sigma^2+\sum_{l=1}^{L}\frac{\beta_{l,i}}{1-\beta_{l,i}}\rho_{l,i}
    }
    \right).
    \label{eq:surrogate_sum_rate_final}
\end{align}

\section{Problem Formulation}

We optimize the transmit powers $\{p_i\}_{i=1}^{d}$ and the distortion factors $\{\beta_{l,i}\}_{l=1,i=1}^{L,d}$ in order to maximize the surrogate sum rate in \eqref{eq:surrogate_sum_rate_final} under the total transmit power constraint
\begin{align}
    \sum_{i=1}^{d}p_i\leq P,
    \quad
    p_i\geq 0,\ \forall i.
\end{align}

Using the high-resolution relation between the quantization bits $b_{l,i}$ and the distortion factors \cite{quantization}, we have $\beta_{l,i}\approx c_q2^{-2b_{l,i}}$ for some multiplicative factor $c_q>0$ depending on the source distribution and quantizer design. The fronthaul constraint can then be written as
\begin{align}
    \sum_{l=1}^{L}\sum_{i=1}^{d}
    \frac{1}{2}\log_2\!\left(\frac{c_q}{\beta_{l,i}}\right)
    \leq b_{\rm tot},
    \quad
    0<\beta_{l,i}\leq c_q,
    \label{eq:fronthaul_constraint_beta}
\end{align}
where $b_{\rm tot}$ is the total number of fronthaul bits available per channel use.

During the continuous optimization phase, all AP--stream branches are assumed to be assigned at least a non-zero resolution, i.e., $\beta_{l,i}\in(0,c_q]$. The upper bound $c_q$ follows from the high-resolution quantization model and may exceed one depending on the quantizer design (e.g., Lloyd--Max quantizers for Gaussian sources). In our formulation, we retain this looser upper bound and empirically observe that it leads to stable and accurate approximations within the proposed framework.

Overall, the optimization problem is formulated as
\begin{subequations}\label{eq:main_problem_current}
\begin{align}
\max_{\{p_i\},\{\beta_{l,i}\}}
\quad &
\sum_{i=1}^{d}
\log_2\left(
1+\frac{p_i\lambda_i^2}{
\sigma^2+\sum_{l=1}^{L}\frac{\beta_{l,i}}{1-\beta_{l,i}}\rho_{l,i}
}
\right)
\\
\text{s.t.}\quad
&
\sum_{i=1}^{d}p_i\leq P,
\\
&
\sum_{l=1}^{L}\sum_{i=1}^{d}
\frac{1}{2}\log_2\!\left(\frac{c_q}{\beta_{l,i}}\right)
\leq b_{\rm tot},
\\
&
p_i\geq 0,\ \forall i,
\quad
0<\beta_{l,i}\leq c_q,\ \forall l,i.
\end{align}
\end{subequations}

The problem in \eqref{eq:main_problem_current} is highly non-convex. To handle this, we adopt WMMSE-based block coordinate descent.

\section{WMMSE-Based Block Coordinate Descent}

Based on \eqref{eq:nui}, the approximate SINR of stream $i$ can be written as $ \bar{\gamma}_i
    =
    \frac{p_i\lambda_i^2}{\nu_i}$. This SINR corresponds to the scalar observation model $
    \tilde{r}_i=\lambda_i\sqrt{p_i}\,s_i+\tilde{z}_i,$
where $\mathbb{E}\left\{|\tilde{z}_i|^2\right\}=\nu_i$.

We let $u_i\in\mathbb{C}$ denote a scalar equalizer and define the mean-squared error (MSE) as
\begin{align}
    e_i
    &=
    \mathbb{E}\left\{|u_i\tilde{r}_i-s_i|^2\right\} \nonumber\\
    &=
    |u_i|^2\left(p_i\lambda_i^2+\nu_i\right)
    -2\Re\!\left\{u_i\lambda_i\sqrt{p_i}\right\}
    +1.
\end{align}
Using the standard WMMSE equivalence, the surrogate sum-rate maximization is equivalent to
\begin{subequations}\label{eq:wmmse_bcd_master}
\begin{align}
\min_{\{p_i\},\{\beta_{l,i}\},\{u_i\},\{w_i\}}
\quad &
\sum_{i=1}^{d}\left(w_i e_i-\log(w_i)\right)
\\
\text{s.t.}\quad
&
\sum_{i=1}^{d}p_i\leq P,
\\
&
\sum_{l=1}^{L}\sum_{i=1}^{d}
\frac{1}{2}\log_2\!\left(\frac{c_q}{\beta_{l,i}}\right)\leq b_{\rm tot},
\\
&
p_i\geq 0,\ \forall i,
\quad
0<\beta_{l,i}\leq c_q,\ \forall l,i,
\\
&
w_i>0,\ \forall i.
\end{align}
\end{subequations}

We solve \eqref{eq:wmmse_bcd_master} via block coordinate descent as follows.

\subsection{Equalizer Update}

For fixed $\{p_i\}$, $\{\beta_{l,i}\}$, and $\{w_i\}$, the optimal equalizer is
\begin{align}
    u_i^\star
    =
    \frac{\lambda_i\sqrt{p_i}}{p_i\lambda_i^2+\nu_i}.
\end{align}

\subsection{Weight Update}

For fixed $\{p_i\}$, $\{\beta_{l,i}\}$, and $\{u_i\}$, the resulting MSE is
\vspace{-2mm}
\begin{align}
    e_i^\star
    =
    1-\frac{p_i\lambda_i^2}{p_i\lambda_i^2+\nu_i},
\end{align}
and the optimal weight update is $ w_i^\star=\frac{1}{e_i^\star}$.

\subsection{Power Update}

For fixed $\{u_i\}$, $\{w_i\}$, and $\{\beta_{l,i}\}$, we define
\begin{align}
    a_{i,k}
    \triangleq
    \sum_{l=1}^{L}
    \frac{\beta_{l,i}}{1-\beta_{l,i}}
    \left|
    \mathbf{u}_{l,i}^{\Htran}\mathbf{H}_l\mathbf{v}_k
    \right|^2.
\end{align}
Then, the power-update subproblem is
\begin{subequations}\label{eq:p_subproblem_correct}
\begin{align}
\min_{\{p_i\}}
\quad &
\sum_{i=1}^{d}
w_i\left(
|u_i|^2
\left(
p_i\lambda_i^2+\sum_{k=1}^{d} a_{i,k}p_k
\right)
-2\Re\{u_i\lambda_i\}\sqrt{p_i}
\right)
\\
\text{s.t.}\quad
&
\sum_{i=1}^{d}p_i\leq P,
\\
&
p_i\geq 0,\ \forall i.
\end{align}
\end{subequations}

By defining $t_i=\sqrt{p_i}$, the problem in \eqref{eq:p_subproblem_correct} can be reformulated as
\begin{subequations}\label{eq:t_subproblem_simplified}
\begin{align}
\min_{\{t_i\}}
\quad &
\sum_{i=1}^{d}
\left(
d_i t_i^2
-2w_i\Re\{u_i\lambda_i\}t_i
\right)
\\
\text{s.t.}\quad
&
\sum_{i=1}^{d}t_i^2\leq P,
\\
&
t_i\geq 0,\ \forall i,
\end{align}
\end{subequations}
where $
    d_i
    =
    w_i|u_i|^2\lambda_i^2
    +
    \sum_{m=1}^{d} w_m |u_m|^2 a_{m,i}$. Via the KKT conditions, we obtain
\begin{align}
    t_i^\star
    =
    \left[
    \frac{w_i\Re\{u_i\lambda_i\}}
    {d_i+\mu}
    \right]^+,
\end{align}
where $\mu\geq 0$ is the Lagrange multiplier associated with the power constraint. 

If the solution obtained with $\mu=0$ satisfies the power constraint, i.e., $\sum_{i=1}^{d}(t_i^\star)^2 \leq P$, then $\mu=0$ is optimal and the constraint is inactive. Otherwise, $\mu>0$ is chosen such that the power constraint is satisfied with equality, i.e., $
    \sum_{i=1}^{d}(t_i^\star)^2 = P$.
Therefore, the optimal power allocation is given by
\begin{align}
    p_i^\star
    =
    \left(
    \left[
    \frac{w_i\Re\{u_i\lambda_i\}}
    {d_i+\mu}
    \right]^+
    \right)^2.
\end{align}

\subsection{$\beta$ Update}

For fixed $\{u_i\}$, $\{w_i\}$, and $\{p_i\}$, the terms depending on $\beta_{l,i}$ are contained in
\vspace{-2mm}
\begin{align}
    \sum_{i=1}^{d}w_i e_i
    =
    \sum_{i=1}^{d}
    w_i
    \left(
    |u_i|^2\nu_i
    -2\Re\{u_i\lambda_i\sqrt{p_i}\}
    \right)
    +\text{const.}
\end{align}
Since only $\nu_i$ depends on $\beta_{l,i}$, substituting its definition yields
\vspace{-2mm}
\begin{align}
    \sum_{i=1}^{d}w_i e_i
    =
    \sum_{l=1}^{L}\sum_{i=1}^{d}
    f_{l,i}\frac{\beta_{l,i}}{1-\beta_{l,i}}
    +\text{const.},
\end{align}
where $f_{l,i}
    \triangleq
    w_i|u_i|^2\rho_{l,i}$. Hence, the $\beta$ update is given by
\vspace{-2mm}\begin{align}
\min_{\{\beta_{l,i}\}}
\quad &
\sum_{l=1}^{L}\sum_{i=1}^{d}
f_{l,i}\frac{\beta_{l,i}}{1-\beta_{l,i}}
\label{eq:beta_bcd_subproblem}
\\
\text{s.t.}\quad
&
\sum_{l=1}^{L}\sum_{i=1}^{d}
\frac{1}{2}\log_2\!\left(\frac{c_q}{\beta_{l,i}}\right)\leq b_{\rm tot},
\\
&
0<\beta_{l,i}\leq c_q,\ \forall l,i.
\end{align}

This problem admits a semi-closed-form solution.\footnote{Here, we implicitly assume that $c_q<1$, but during algorithmic implementation, it is allowed to take larger values than one.} By introducing the Lagrange multiplier $\eta\ge 0$ for the fronthaul constraint, the KKT conditions yield
\vspace{-2mm}
\begin{align}
    \frac{f_{l,i}}{(1-\beta_{l,i})^2}
    -\frac{\eta}{2\ln 2}\frac{1}{\beta_{l,i}}=0.
\end{align}
For $f_{l,i}>0$, the above equation is equivalent to
\begin{align}
    f_{l,i}\beta_{l,i}
    =
    \frac{\eta}{2\ln 2}(1-\beta_{l,i})^2,
\end{align}
which is a quadratic equation in $\beta_{l,i}$. Therefore, the solution can be written as
\vspace{-2mm}
\begin{align}
    \beta_{l,i}^\star
    =
    \min\!\left\{
    c_q,\,
    \bar{\beta}_{l,i}(\eta)
    \right\},
\end{align}
where $\bar{\beta}_{l,i}(\eta)$ denotes the valid root of
\vspace{-2mm}
\begin{align}
    \frac{\eta}{2\ln 2}(1-\beta)^2-f_{l,i}\beta=0.
\end{align}
The scalar multiplier $\eta$ is selected such that the fronthaul constraint is satisfied with equality.

During the WMMSE iterations, AP-stream branches with large distortion factors are treated as effectively inactive in the distortion modeling. Specifically, when $\beta_{l,i}$ exceeds a predefined threshold, the corresponding distortion contribution $\frac{\beta_{l,i}}{1-\beta_{l,i}}$ is omitted. After convergence, the resulting continuous distortion factors are projected onto discrete quantization levels. If no bit is assigned to a given AP-stream branch after this projection, that branch is deactivated and no quantized signal is forwarded over the fronthaul.

\vspace{-2mm}

\section{Numerical Results}

In this section, we evaluate the performance of the proposed WMMSE-based joint power and resolution allocation scheme and compare it with two baseline strategies. We consider the uplink of a cell-free massive MIMO system deployed over a square area of $250\times 250$ meters. A total of $L=25$ APs are randomly distributed in the coverage area, each equipped with $N=4$ antennas. The vertical distance between the APs and the UE is set to $10$ meters. The path loss in decibel scale is given according to Urban Microcell Street Canyon model as $-32.4-20\log_{10}(f_c)-31.9\log_{10}(d)$ where $d$ is the distance between two nodes in meters, $f_c=3.5$\,GHz is the carrier frequency \cite[Table 7.4.1-1]{3GPP5G}. The small-scale fading channel coefficients are generated under uncorrelated Rayleigh fading. The bandwidth is $50$\,MHz and the noise figure is $5$\,dB. The total UE transmit power is $P=1$\,W. and the high-resolution quantization constant is set to $c_q = \pi\sqrt{3}/2$, corresponding to a Lloyd--Max quantizer designed for Gaussian signals. 

For each channel realization, we evaluate the following three schemes:
\begin{itemize}
    \item \textbf{Proposed WMMSE-based scheme:} Joint optimization of transmit powers and per-AP per-stream quantization resolutions using the proposed WMMSE algorithm.
    \item \textbf{Uniform allocation:} Equal bit allocation across all APs and streams, with equal power allocation across streams.
    \item \textbf{AP-proportional allocation:} Bits are allocated equally across streams, while within each stream the bits are distributed across APs proportionally to the effective channel gains.
\end{itemize}

The performance of all the schemes is evaluated using the exact achievable rate in \eqref{eq:Rexact}, where the covariance matrix of the effective disturbance is estimated via Monte-Carlo simulations with Lloyd--Max quantization applied to each AP-stream branch.

\begin{figure}[t] 
    \centering
        \includegraphics[width=0.45\textwidth, trim=0.8cm 0.2cm 1cm 0.2cm, clip]{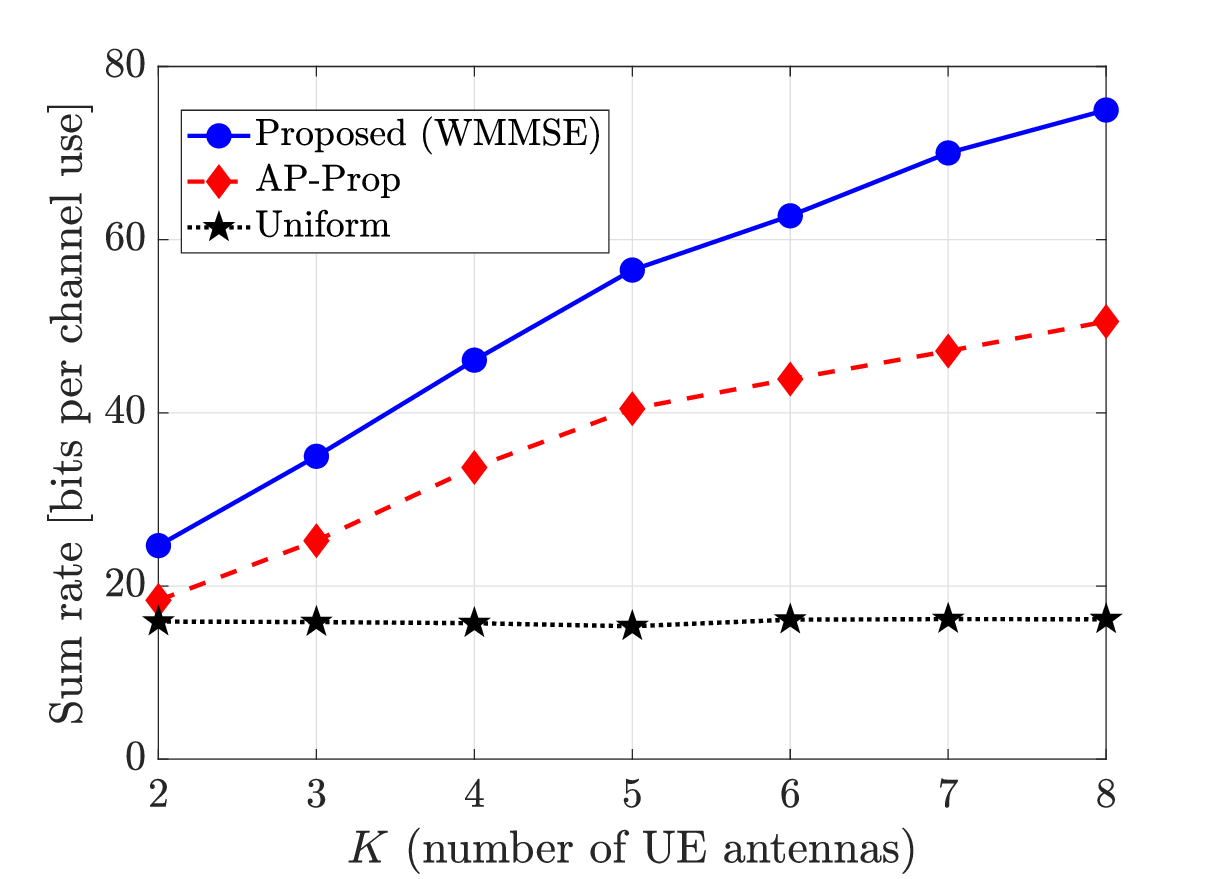} 
         \vspace{-1mm}
        \caption{Sum rate versus the number of UE antennas, $K$, for $b_{\rm tot}=200$\,bits. }
        \label{fig1}
          \vspace{-5mm}
 \end{figure}

Fig.~\ref{fig1} illustrates the sum rate performance as a function of the number of UE antennas $K$, where the total fronthaul budget is fixed to $b_{\rm tot}=200$. The uniform allocation scheme remains almost insensitive to $K$. In contrast, both the proposed WMMSE-based scheme and the AP-proportional allocation significantly benefit from increasing $K$, as they adapt the bit allocation to the underlying channel structure. Moreover, the performance gap between the proposed scheme and the AP-proportional method widens with $K$, highlighting the advantage of joint optimization in effectively utilizing the increased spatial multiplexing capability.

\begin{figure}[t] 
    \centering
        \includegraphics[width=0.45\textwidth, trim=0.8cm 0.2cm 1cm 0.2cm, clip]{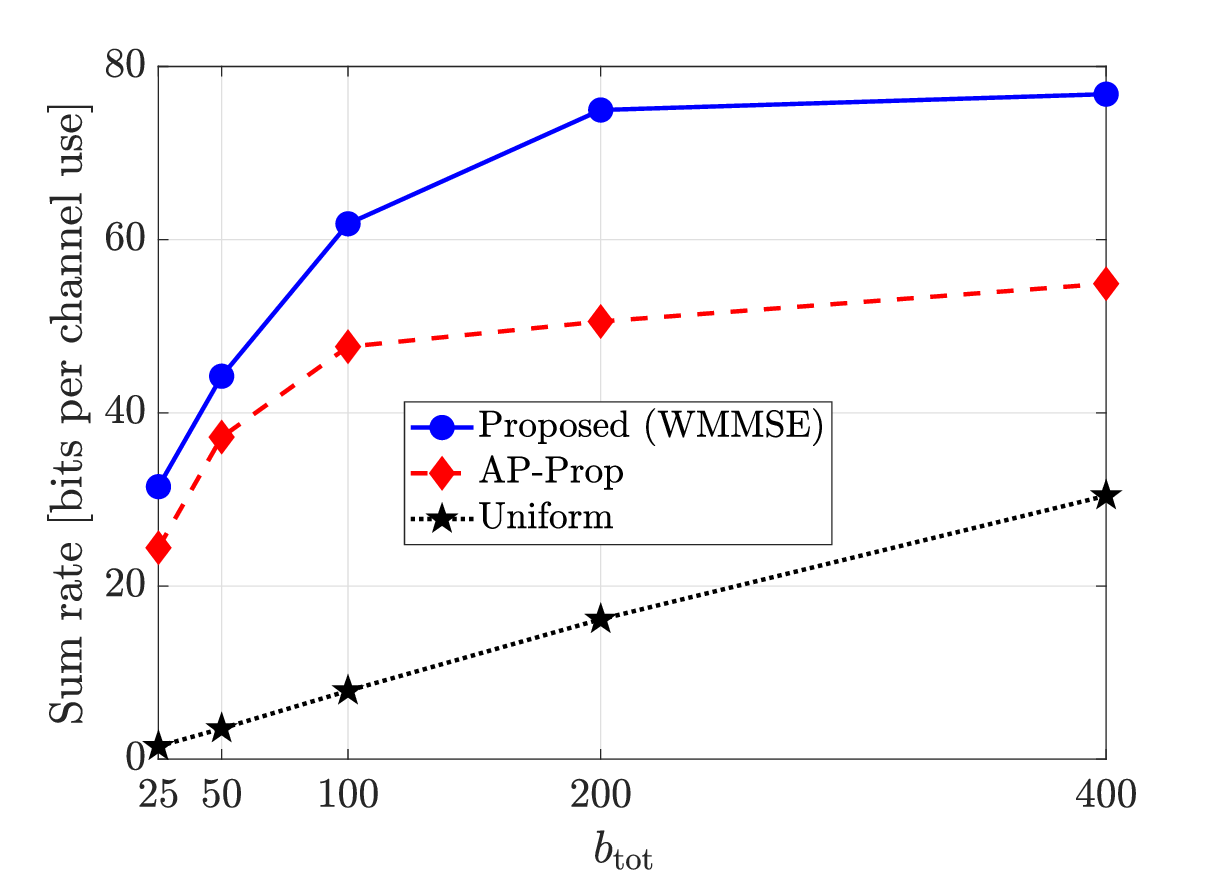} 
         \vspace{-1mm}
        \caption{Sum rate versus $b_{\rm tot}$ for $K=8$. }
        \label{fig2}
          \vspace{-6mm}
 \end{figure}

Fig.~\ref{fig2} shows the sum rate performance as a function of the total fronthaul budget $b_{\rm tot}$ for $K=8$. As expected, all the schemes benefit from increasing fronthaul capacity, since more quantization bits reduce distortion and improve signal reconstruction at the CPU. However, the proposed WMMSE-based scheme consistently achieves the highest sum rate across all operating points, as it jointly optimizes power and bit allocation. The AP-proportional scheme also benefits from additional bits but remains suboptimal due to its heuristic allocation strategy.

\section{Conclusion}

We investigated the uplink of fronthaul-constrained cell-free massive MIMO systems with a multi-antenna UE and proposed a stream-adaptive resolution control framework. Using a distributed SVD-based processing structure, we facilitated per-AP per-stream quantization and formulated a joint optimization of transmit power and quantization distortion under a fronthaul constraint. The resulting problem was efficiently solved via a WMMSE-based algorithm with semi-closed-form updates. Numerical results demonstrated that the proposed approach significantly outperforms conventional uniform and heuristic schemes, particularly by effectively exploiting spatial multiplexing gains under limited fronthaul resources.

\bibliographystyle{IEEEtran}

\bibliography{IEEEabrv,refs}

\end{document}